\documentclass[reqno]{amsart}

\usepackage{booktabs} 
\usepackage{IEEEtrantools}
\usepackage{amssymb,latexsym,amsfonts,amsmath}
\usepackage{graphicx}
\usepackage{mathrsfs}
\usepackage{dsfont}

\usepackage{tikz}
\usetikzlibrary{calc,shapes,arrows}

\usepackage{algorithmic}

\usepackage{xspace}

\usepackage{booktabs} 
\usepackage{latexsym,amsfonts,amsmath}
\usepackage{hyperref}
\usepackage{subfigure}
\usepackage{dsfont}
\usepackage{extarrows}
\usepackage[ruled,vlined,linesnumbered]{algorithm2e}

\usepackage{paralist}
\usepackage{capt-of}
\usepackage{xcolor}

\newcommand{\B}{{\mathbb B}}

\definecolor{myco}{rgb}{0.55, 0.0, 0.63}

\newcommand\blfootnote[1]{%
	\begingroup
	\renewcommand\thefootnote{}\footnote{#1}%
	\addtocounter{footnote}{-1}%
	\endgroup
}

\newtheorem{theorem}{Theorem}[section]
\newtheorem{lemma}[theorem]{Lemma}
\newtheorem{problem}[theorem]{Problem}

\newtheorem{corollary}[theorem]{Corollary}

\newtheorem{definition}[theorem]{Definition}

\newtheorem{remark}[theorem]{Remark}

\numberwithin{equation}{section}

\begin{document}
	
\begin{abstract}
In this paper, we provide a direct data-driven approach to synthesize safety controllers for unknown linear systems affected by unknown-but-bounded disturbances, in which identifying the unknown model is not required. 
First, we propose a notion of $\gamma$-robust safety invariant ($\gamma$-RSI) sets and their associated state-feedback controllers, which can be applied to enforce invariance properties. 
Then, we formulate a data-driven computation of these sets in terms of convex optimization problems with linear matrix inequalities (LMI) as constraints, which can be solved based on a finite number of data collected from a single input-state trajectory of the system.
To show the effectiveness of the proposed approach, we apply our results to a 4-dimensional inverted pendulum.
\end{abstract}

\title[Data-driven Synthesis of Safety Controllers for Uncertain Linear Systems]{Synthesizing Safety Controllers for Uncertain Linear Systems: A Direct Data-driven Approach}

\author{Bingzhuo Zhong$^{1*}$}
\author{Majid Zamani$^{2,3}$}
\author{Marco Caccamo$^{1}$}
\blfootnote{*Corresponding Author.}
\address{$^1$TUM School of Engineering and Design, Technical University of Munich, Germany.}
\email{bingzhuo.zhong@tum.de}
\email{mcaccamo@tum.de}
\address{$^2$Department of Computer Science, University of Colorado Boulder, USA.}
\email{majid.zamani@colorado.edu}
\address{$^3$Department of Computer Science, LMU Munich, Germany.}
\maketitle

\section{Introduction}\label{sec1}
Ensuring the safety of control systems has received significant attentions in the past two decades due to the increasing number of safety-critical real-life applications, such as unmanned aerial vehicles and autonomous transportations.
When models of these applications are available, various model-based techniques can be applied for synthesizing safety controllers, see e.g.,~\cite{Reissig2017Feedback,Ames2016Control,Rungger2017Computing}, to name a few.
Nevertheless, obtaining an accurate model requires a significant amount of effort~\cite{Hou2013model}, and even if a model is available, it may be too complex to be of any use.
Such difficulties motivate researchers to enter the realm of data-driven control methods.
In this paper, we focus on data-driven methods for constructing safety controllers, which enforce invariance properties over unknown linear systems affected by disturbances (i.e., systems are expected to stay within a safe set). 

In general, data-driven control methods can be classified into indirect and direct approaches.
\emph{Indirect data-driven approaches} consist of a system identification phase followed by a model-based controller synthesis scheme.
To achieve a rigorous safety guarantee, it is crucial to provide an upper bound for the error between the identified model and the real but unknown model (a.k.a. \emph{identification error}).
Among different system identification approaches,~\emph{least-squares methods} (see e.g.~\cite{Ljung1999System}) are frequently used for identifying linear models.
In this case, \emph{sharp error bounds}~\cite{Simchowitz2018Learning} relate the identification error to the cardinality of the finite data set which is used for the identification task.
Computation of such bounds requires knowledge about the distributions of the disturbances (typically i.i.d. Gaussian or sub-Gaussian, see e.g.~\cite{Matni2019tutorial,Matni2019self}, and references herein).
Therefore, computation of these bounds is challenging when dealing with~\emph{unknown-but-bounded} disturbances~\cite{Bisoffi2021Trade}, i.e., the disturbances are only assumed to be contained within a given bounded set, but their distributions are fully unknown.  
Note that~\emph{set-membership identification approaches}~(see e.g.~\cite{Lauricella2020Set,Cerone2016Mimo}) can be applied to identify linear control systems with unknown-but-bounded disturbances.
Nevertheless, it is still an open problem to provide an upper bound for the identification error when unknown-but-bounded disturbances are involved.

Different from indirect data-driven approaches,~\emph{direct data-driven approaches} directly map data into the controller parameters without any intermediate identification phase.
Considering systems without being affected by exogenous disturbances, results in~\cite{DePersis2019Formulas} propose a data-driven framework to solve linear quadratic regulation (LQR) problems for linear systems.
Later on, similar ideas were utilized to design model-reference controllers (see~\cite[Section 2]{Breschi2021Direct}) for linear systems~\cite{Breschi2021Direct}, and to stabilize polynomial systems~\cite{Guo2021Data}, switched linear systems~\cite{Rotulo2021Online}, and linear time-varying systems~\cite{Nortmann2020Data}.
When exogenous disturbances are also involved in the system dynamics, recent results, e.g.,~\cite{DePersis2021Low,Berberich2020Robust,Berberich2020Combining,Waarde2020noisy}, can be applied to LQR problems and robust controller design. 
However, none of these results considers state and input constraints.
Hence, they cannot be leveraged to enforce invariance properties.
When input constraints are considered, results in~\cite{Bisoffi2020Data,Bisoffi2020Controller} provide data-driven approaches for constructing state-feedback controllers to make a given \emph{C-polytope} (i.e., \emph{compact} polyhedral set containing the origin~\cite[Definition 3.10]{Blanchini2015Set}) robustly invariant (see~\cite[Problem 1]{Bisoffi2020Controller}).
However, when such controllers do not exist for the given $C$-polytope, one may still be able to find controllers making a subset of this polytope robustly invariant, which is not considered in~\cite{Bisoffi2020Data,Bisoffi2020Controller}.
Additionally, the approaches in~\cite{Bisoffi2020Data,Bisoffi2020Controller} require an individual constraint for each vertex of the polytope (see~\cite[Section 4]{Bisoffi2020Data} and~\cite[Theorem 1 and 2]{Bisoffi2020Controller}). 
Unfortunately, given any arbitrary polytope, the number of its vertices grows exponentially with respect to its dimension and the number of hyperplanes defining it in the worst case~\cite[Section 1]{Dyer1983complexity}.

In this paper, we focus on enforcing invariance properties over unknown linear systems affected by unknown-but-bounded disturbances.
Particularly, we propose a direct data-driven approach for designing safety controllers against these properties.
To this end, we first propose so-called $\gamma$-robust safety invariant ($\gamma$-RSI) sets and their associated state-feedback controllers enforcing invariance properties modeled by (possibly \emph{unbounded}) polyhedral safety sets.
Then, we propose a data-driven approach for computing such sets, in which the numbers of constraints and optimization variables grow linearly with respect to the numbers of hyperplanes defining the safety set and the cardinality of the finite data set.
Moreover, we also discuss the relation between our data-driven approach and the condition of~\emph{persistency of excitation}~\cite{Willems2005note}, which is a crucial concept in most literature about direct data-driven approaches.

The remainder of this paper is structured as follows.
In Section~\ref{sec2}, we provide preliminary discussions on notations, models, and the underlying problems to be tackled.
Then, we propose in Section~\ref{sec3} the main results for the data-driven approach.
Finally, we apply our methods to a 4-dimensional inverted pendulum in Section~\ref{sec4} and conclude our results in Section~\ref{sec5}.
For a streamlined presentation, the proofs of all results in this paper are provided in the Appendix.

\section{Preliminaries and Problem Formulation}\label{sec2}
\subsection{Notations}
We use $\mathbb{R}$ and $\mathbb{N}$ to denote the sets of real and natural numbers, respectively. 
These symbols are annotated with subscripts to restrict the sets in a usual way, e.g., $\mathbb{R}_{\geq0}$ denotes the set of non-negative real numbers.
Moreover, $\mathbb{R}^{n\times m}$ with $n,m\in \mathbb{N}_{\geq 1}$ denotes the vector space of real matrices with $n$ rows and $m$ columns.
For $a,b\in\mathbb{R}$ (resp. $a,b\in\mathbb{N}$) with $a\leq b$, the closed, open and half-open intervals in $\mathbb{R}$ (resp. $\mathbb{N}$) are denoted by $[a,b]$, $(a,b)$ ,$[a,b)$ and $(a,b]$, respectively. 
We denote by $\mathbf{0}_{n\times m}$ and $\mathbf{I}_n$ the zero matrix in $\mathbb{R}^{n\times m}$, and the identity matrix in $\mathbb{R}^{n\times n}$, respectively. 
Their indices are omitted if the dimension is clear from the context.
Given $N$ vectors $x_i \in \mathbb R^{n_i}$, $n_i\in \mathbb N_{\ge 1}$, and $i\in\{1,\ldots,N\}$, we use $x = [x_1;\ldots;x_N]$ to denote the corresponding column vector of the dimension $\sum_i n_i$.
Given a matrix $M$, 
we denote by $\text{rank}(M)$, $\text{det}(M)$, $M^\top$, $M(i)$, and  $M(i,j)$, the rank, the determinant, the transpose, the $i$-th column, and the entry in $i$-th row and $j$-th column of $M$, respectively.

\subsection{System}
In this paper, we focus on discrete-time linear control systems defined as
\begin{equation}\label{eq:linear_subsys}
x(k+1) = Ax(k) +Bu(k) + d(k), \quad k\in\mathbb N,
\end{equation}
with $A\in\mathbb{R}^{n\times n}$ and $B\in\mathbb{R}^{n\times m}$ being some unknown constant matrices; $x(k)\in X$ and $u(k)\in U$, $\forall k\in\mathbb{N}$, being the state and the input vectors, respectively, in which $X\subseteq \mathbb{R}^n$ is the state set,
\begin{align}
U = \{u \in \mathbb{R}^m|b_ju\leq 1,j = 1,\ldots,\mathsf{j}\}\subset \mathbb{R}^m, \label{input_set}
\end{align}
is the input set of the system, with $b_j\in \mathbb{R}^m$ being some known vectors; 
$d(k)$ denotes the exogenous disturbances, where $d(k)\in \Delta(\gamma)$, $\forall k\in \mathbb{N}$, with
\begin{align}
\Delta(\gamma) = \{d\in\mathbb{R}^n | d^\top d\leq\gamma,\gamma\in \mathbb{R}_{\geq 0} \}\label{eq:disturbance_set}.
\end{align}
Note that disturbances of the form of~\eqref{eq:disturbance_set} are also known as~\emph{unknown-but-bounded disturbance with instantaneous constraint}~\cite{Bisoffi2021Trade}, with $\gamma$ being the disturbance bound that is assumed to be a priori.
Finally, we denote by
\begin{align}
X_{1,N} &:= \begin{bmatrix}x(1)&x(2)&\ldots&x(N)\end{bmatrix},\label{eq:state_seq1}\\
X_{0,N} &:= \begin{bmatrix}x(0)&x(1)&\ldots&x(N-1)\end{bmatrix},\label{eq:state_seq}\\
U_{0,N}&:=\begin{bmatrix}u(0) &u(1)  & \ldots & u(N-1)\end{bmatrix},\label{eq:inputseqm}
\end{align}
the data collected offline, with $N\!\in\! \mathbb{N}$, in which $x(0)$ and $U_{0,N}$ are chosen by the users, while the rest are obtained by observing the state sequence generated by the system in~\eqref{eq:linear_subsys}.

\subsection{Problem Formulation}
In this paper, we are interested in invariance properties, which can be modeled by (possibly unbounded) safety sets defined as
\begin{align}
S := \{x\in\mathbb{R}^n| a_ix\leq 1, i = 1,\ldots,\mathsf{i}\}\subset X\label{safety_set},
\end{align}
where $a_i\in \mathbb{R}^{n}$ are some known vectors.
The main problem in this paper is formulated as follows.
\begin{problem}\label{prob}
	Consider a linear control system as in~\eqref{eq:linear_subsys}, where matrices $A$ and $B$ are unknown, with input set as in~\eqref{input_set}, and safety set as in~\eqref{safety_set}.
	Using data in~\eqref{eq:state_seq1}-~\eqref{eq:inputseqm}, design a \emph{safety envelope} $\bar{\mathcal{S}}\subseteq S$ along with a \emph{safety controller} $u=Kx$ (if existing) such that $x(k)\in \bar{\mathcal{S}}$, $\forall k\in\mathbb{N}_{>0}$, if $x(0)\in\bar{\mathcal{S}}$.
\end{problem}

\section{Main Results}\label{sec3}
\subsection{$\gamma$-Robust Safety Invariant Set}
In this subsection, we propose the computation of $\gamma$-robust safety invariant ($\gamma$-RSI) sets assuming matrices $A$ and $B$ in~\eqref{eq:linear_subsys} are known.
These sets would be later employed as safety envelopes as defined in Problem~\ref{prob}.
Then, we utilize these results in the next subsection to provide the main direct data-driven approach to solve Problem~\ref{prob}.
First, we present the definition of $\gamma$-RSI sets as follows.
\begin{definition}\label{def:RSI}
	($\gamma$-RSI set)
	Consider a linear control system as in~\eqref{eq:linear_subsys}.
	A $\gamma$-RSI set $\mathcal{S}$ with respect to a safety set $S$ as in~\eqref{safety_set} is defined as
	\begin{equation}
	\mathcal{S}:=\{x\in\mathbb{R}^n|x^\top Px\leq 1 \}\subset S,\label{eq:safety_set}
	\end{equation} 
	such that $\forall x\in \mathcal{S}$, one has $Ax +Bu + d\in \mathcal{S}$, $\forall d\in\Delta(\gamma)$, when the \emph{RSI-based controller}
	\begin{equation}
	u=Kx,\label{eq:safety_controller}
	\end{equation}
	associated with $\mathcal{S}$ is applied in the closed-loop, where $P\in \mathbb{R}^{n \times n}$ is a positive-definite matrix, and $K\in\mathbb{R}^{m\times n}$.
\end{definition}

With this definition, we present the next straightforward result for Problem~\ref{prob}, which can readily been verified according to Definition~\ref{def:RSI}.
\begin{theorem}\label{thm:solveprob}
	Consider a system as in~\eqref{eq:linear_subsys}.
	If there exists a $\gamma$-RSI set $\mathcal{S}$ as in~\eqref{eq:safety_set}, then one has $x(k)\in \mathcal{S}$, $\forall k \!\in\!\mathbb{N}_{>0}$, when the RSI-based controller as in~\eqref{eq:safety_controller} associated with $\mathcal{S}$ is applied in the closed-loop, and $x(0)\in\mathcal{S}$.
\end{theorem}
\begin{remark}
	In this work, we focus on computing elliptical-type $\gamma$-RSI sets to solve Problem~\ref{prob}, while computing $\gamma$-RSI sets of more general forms, e.g., polyhedral-type sets, is left to future investigations.
	One of the difficulties of computing polyhedral-type $\gamma$-RSI sets is to cast the volume of a polyhedral set as a convex objective function~\cite[Section 2]{Khachiyan1993Complexity}, which is done easily in the elliptical case (cf. Remark~\ref{objective}).
	Additionally, consider an $\mathsf{n}$-dimensional polytope $\mathcal{P}\subseteq \mathbb{R}^{\mathsf{n}}$, which is defined by $\mathsf{m}$ hyperplanes.
	The model-based approaches (see e.g.~\cite{Blanchini1990Feedback}) require an individual constraint for each vertex of $\mathcal{P}$ for synthesizing controllers that make $\mathcal{P}$ a $\gamma$-RSI set. 
	Therefore, we suspect that the exponential growth in the number of vertices with respect to $\mathsf{n}$ and $\mathsf{m}$~\cite[Section 1]{Dyer1983complexity} could also be a burden for extending our data-driven approach to polyhedral-type $\gamma$-RSI sets.
\end{remark}

Using Theorem~\ref{thm:solveprob}, the other question is how to compute $\gamma$-RSI sets.
To do so, we need the following result.
\begin{theorem}\label{thm:LMI_1}
	Consider a system as in~\eqref{eq:linear_subsys}. 
	For any matrix $K\in\mathbb{R}^{m\times n}$, positive-definite matrix $P\in \mathbb{R}^{n \times n}$, and $\gamma\in\mathbb{R}_{\geq 0}$, one has
	\begin{align}
	\Big((A+BK)x+d\Big)^\top P\Big((A+BK)x+d\Big)\leq 1,\label{ineq:safety_set}
	\end{align}
	$\forall d\in\Delta(\gamma)$, and $\forall x\in \mathbb{R}^n$ satisfying $ x^\top Px\leq 1$, if and only if $\exists \kappa \in (0,1]$, such that
	\begin{enumerate}
		\item (\textbf{Cond.1}) $x^\top (A+BK)^\top P(A+BK)x\leq \kappa$ holds $\forall x\in \mathbb{R}^n$ satisfying $x^\top Px\leq 1$;
		\item (\textbf{Cond.2}) $(y+\tilde{d})^\top P(y+\tilde{d})\leq 1$ holds $\forall y\in \mathbb{R}^n$ satisfying $y^\top Py\leq \kappa$, and $\forall \tilde{d}\in\Delta(\gamma)$.
	\end{enumerate} 
\end{theorem}

The proof of Theorem~\ref{thm:LMI_1} is provided in the Appendix. 
In Figure~\ref{fig1:intuition}, we provide some intuitions for Theorem~\ref{thm:LMI_1}.
\begin{figure}
	\centering
	\includegraphics[width=4.2cm]{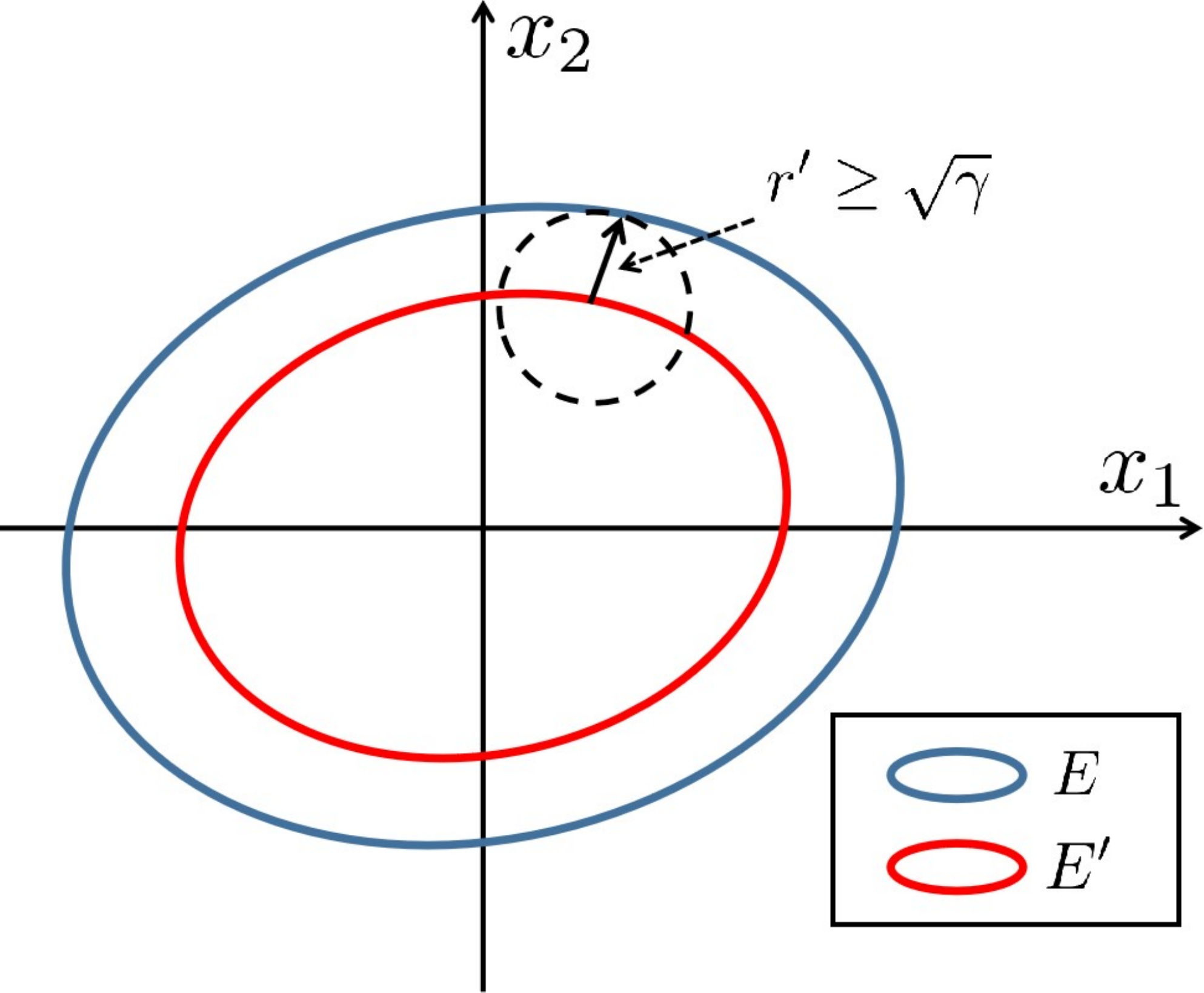}
	\caption{An envelope $E:=\{x\in \mathbb{R}^n|x^\top Px\leq 1\}$ is a $\gamma$-RSI set, when there exists a controller $u=Kx$ that can steer any $x \in E$ into a smaller envelope $E':=\{x^+\in \mathbb{R}^n|(x^+)^\top Px^+\leq \kappa\}$ in which we assume $d=\mathbf{0}$, i.e., $\forall x\in E$, one gets $x^+\in E'$, with $x^+=(A+BK)x$.} \label{fig1:intuition}
\end{figure}
Next, we propose an optimization problem for computing a $\gamma$-RSI set for a linear control system as in~\eqref{eq:linear_subsys}, assuming that matrices $A$ and $B$ are known.

\begin{definition}\label{opt1}
	Consider a linear system as in~\eqref{eq:linear_subsys} with input constraints in~\eqref{input_set}, a safety set $S$ in~\eqref{safety_set}, $\kappa\in(0,1]$, and $\gamma\geq 0$.
	We define an optimization problem, denoted by $OP_m$ as:
	\begin{align}
	OP_m: \min_{Q,\bar{K}} &-\text{log}(\text{det}(Q))~\label{eq:objective_function}\\
	\mbox{s.t.}\ &\begin{bmatrix}\kappa Q & Q^\top A^\top +\bar{K}^\top B^\top \\AQ+B\bar{K} &Q\end{bmatrix}\succeq 0,\label{synab_cond1}
	\end{align}
	\begin{align}
	&Q\succeq c\mathbf{I},\label{synab_cond2}\\
	&a_iQa^\top _i\leq 1,\,i=1,\ldots, \textsf{i},\label{synab_cond3}\\
	&\begin{bmatrix}1 & b_j\bar{K}\\\bar{K}^\top b_j^\top  &Q\end{bmatrix}\succeq 0,\,j=1,\ldots, \textsf{j}\label{synab_cond4},
	\end{align}
	where $c=\frac{\gamma}{(1-\sqrt{\kappa})^2}$ if $\kappa \neq 1$, and $c=0$ otherwise; $Q\in \mathbb{R}^{n \times n}$ is a positive-definite matrix, and $\bar{K}\in\mathbb{R}^{m\times n}$.
\end{definition}

Based on Definition~\ref{opt1}, one can construct an RSI-based controller enforcing invariance properties as in the next result.
\begin{theorem}\label{cor:synab}
	Consider the optimization problem $OP_m$ in Definition~\ref{opt1}.
	For any $\kappa\in(0,1]$ and $\gamma\geq 0$, the set $\mathcal{S}':=\{x\in X|x^\top Q^{-1}x\leq 1\}$ is a $\gamma$-RSI set with $u = \bar{K}Q^{-1}x$ being the associated RSI-based controller, if and only if $OP_m$ is feasible for the given $\gamma$ and $\kappa$. 
\end{theorem}

The proof for Theorem~\ref{cor:synab} can be found in the Appendix.
Note that the existence of $\kappa\in(0,1]$ is a necessary and sufficient condition for the existence of a $\gamma$-RSI set with respect to the safety set $S$ as in~\eqref{safety_set} according to Theorem~\ref{thm:LMI_1}.
In practice, one can apply bisection to come up with the largest value of $\kappa$ while solving $OP_m$.
\begin{remark}\label{objective}
	The objective function in~\eqref{eq:objective_function} maximizes the volume of the $\gamma$-RSI set in Theorem~\ref{cor:synab}, since its volume is proportional to $\text{det}(Q)$~\cite[p. 42]{Boyd1994Linear}.
\end{remark}

So far, we have proposed an approach for computing $\gamma$-RSI sets by assuming matrices $A$ and $B$ are known.
Before proposing the direct data-driven approach with the help of the results in this subsection, we want to point out the challenge in solving Problem~\ref{prob} using indirect data-driven approaches.	
Following the idea of indirect data-driven approaches, one needs to identify unknown matrices $A$ and $B$ based on data, and then applies Theorem~\ref{cor:synab} to the identified model 
\begin{equation*}
x(k+1)= \hat{A}x(k)+\hat{B}u(k)+\hat{d}(k),
\end{equation*}
where $\hat{A}$ and $\hat{B}$ are the estimation of $A$ and $B$, respectively, and $\hat{d}(k) := (A - \hat{A})x(k) + (B - \hat{B}) u(k) +d(k)$, with $d(k)\in\Delta(\gamma)$.
Accordingly, one has $\lVert \hat{d}(k)\rVert\leq \Delta_A \lVert x(k)\rVert+\Delta_B\lVert u(k)\rVert+\gamma$, with $\Delta_A\! :=\! \lVert A - \hat{A} \rVert$ and $\Delta_B \!:=\!\lVert B - \hat{B} \rVert$.
Here, $\Delta_A$ and $\Delta_B$ are known as \emph{sharp error bounds}~\cite{Simchowitz2018Learning}, which relate the identification error to the cardinality of the finite data set used for system identification.	Note that the computation of these bounds requires some assumptions on the distribution of the disturbances (typically disturbances with symmetric density functions around the origin such as Gaussian and sub-Gaussian, see discussion in  e.g.~\cite{Matni2019tutorial,Matni2019self} and references herein).
To the best of our knowledge, it is still an open problem how to compute such bounds when considering unknown-but-bounded disturbances (also see the discussion in Section~\ref{sec1}).
Such challenges in leveraging indirect data-driven approaches motivated us to propose a direct data-driven approach for computing $\gamma$-RSI sets, in which the intermediate system identification step is not required.

\subsection{Direct Data-driven Computation of $\gamma$-RSI Sets}\label{sec4.1}
In this subsection, we propose a direct data-driven approach for computing $\gamma$-RSI sets.
To this end, the following definition is required.
\begin{definition}\label{opt3}
	Consider a linear system as in~\eqref{eq:linear_subsys} with input constraints as in~\eqref{input_set}, a safety set $S$ as in~\eqref{safety_set}, $X_{1,N}$, $X_{0,N}$, and $U_{0,N}$, as in~\eqref{eq:state_seq1}-\eqref{eq:inputseqm}, respectively.
	Given $\kappa\in(0,1]$ and $\gamma\geq 0$, we define an optimization problem, denoted by $OP_d$ as:
	\begin{align}
	OP_d:\!\!\!\!\!\!\min_{Q,\bar{Z},\epsilon_1,\ldots,\epsilon_N}\!\!\!\!\! &-\text{log}(\text{det}(Q))\label{eq:objective_function2} \\
	\mbox{s.t.}\ 
	&Q\succeq c\mathbf{I},\label{LMI2_cond1}\\
	& N_1\!	-\!\sum_{p=1}^{N}\epsilon_p N_p
	\begin{bmatrix}
	\gamma \mathbf{I}_n    &\mathbf{0} \\
	\mathbf{0}   &-1
	\end{bmatrix}
	N_p^\top \!\succeq\! 0;\label{LMI2_cond2}\\
	&a_iQa^\top _i\leq 1,\,i=1,\ldots, \textsf{i},\label{LMI2_cond3}\\
	&\begin{bmatrix}1 & b_j\bar{Z}\\\bar{Z}^\top b_j^\top  &Q\end{bmatrix}\succeq 0,\,j=1,\ldots, \textsf{j},\label{LMI2_cond4}
	\end{align}
	where $\epsilon_i>0$, $\forall i\in[1,N]$,
	\begin{align*}
	N_1 = \begin{bmatrix}
	\kappa Q    &\mathbf{0}  & \mathbf{0} & \mathbf{0}\\
	\mathbf{0}  &-Q   & -\bar{Z}^\top  & \mathbf{0}\\
	\mathbf{0} 	&-\bar{Z} & \mathbf{0}& \bar{Z} \\
	\mathbf{0}  &\mathbf{0} & \bar{Z}^\top  &Q        
	\end{bmatrix};
	N_p = \begin{bmatrix}
	\mathbf{I}_n  &X_{1,N}(p) \\
	\mathbf{0}  &-X_{0,N}(p)  \\
	\mathbf{0}	&-U_{0,N}(p) \\
	\mathbf{0}  &\mathbf{0}      
	\end{bmatrix},
	\end{align*}
	$\forall p\in[1,N]$; $c=\frac{\gamma}{(1-\sqrt{\kappa})^2}$ if $\kappa \neq 1$, and $c=0$, otherwise;	$Q\in\mathbb{R}^{n\times n}$  is a positive-definite matrix, and $\bar{Z}\in\mathbb{R}^{m\times n}$.
\end{definition}
With the help of Definition~\ref{opt3}, we propose the following result for building an RSI-based controller with respect to invariance properties.
\begin{theorem}\label{thm:imp2}
	Consider an optimization problem $OP_d$ as in Definition~\ref{opt3} and the disturbance set $\Delta(\gamma)$ as in~\eqref{eq:disturbance_set}.
	For any $\kappa\in(0,1]$, if $OP_d$ is feasible, then the set $\mathcal{S}'_d:=\{x\in X|x^\top Q^{-1}x\leq 1\}$ is a $\gamma$-RSI set, with $u=\bar{Z}Q^{-1}x$ being the RSI-based controller associated with $\mathcal{S}'_d$.
\end{theorem}

The proof of Theorem~\ref{thm:imp2} is provided in the Appendix.
It is also worth mentioning that the number of LMI constraints in $OP_d$ grows linearly with respect to the number of inequalities defining the safety set in~\eqref{safety_set} and input set in~\eqref{input_set}. 
Meanwhile, the sizes of the (unknown) matrices on the left-hand sides of~\eqref{LMI2_cond1}-\eqref{LMI2_cond4} are independent of the number of data, i.e., $N$, and grow linear with respect to the dimensions of the state and input sets.
Additionally, the number of slack variables, i.e., $\epsilon_i$, grows linearly with respect to $N$.
As a result, the optimization problem $OP_d$ in Definition~\ref{opt3} can be solved efficiently.

\begin{remark}
	Although in~Theorem~\ref{cor:synab} (assuming matrices $A$ and $B$ are known), the feasibility of $OP_m$ for given $\gamma$ and $\kappa$ is a necessary and sufficient condition for the existence of $\gamma$-RSI sets, Theorem~\ref{thm:imp2} only provides a sufficient condition on the existence of such sets.
	As a future direction, we plan to work on a direct data-driven approach that provides necessary and sufficient conditions for computing $\gamma$-RSI sets, but this is out of the scope of this work.
\end{remark}

In the remainder of this section, we discuss our proposed direct data-driven approach in terms of the condition of \emph{persistency of excitation}~\cite{Willems2005note} regarding the offline-collected data $X_{0,N}$ and $U_{0,N}$.
We first recall this condition, which is adapted from~\cite[Corollary 2]{Willems2005note}.
\begin{lemma}\label{PEassumption}
	Consider the linear system in~\eqref{eq:linear_subsys} with $(A,B)$ being controllable, $X_{0,N}$ as in~\eqref{eq:state_seq}, and $U_{0,N}$ as in~\eqref{eq:inputseqm}.
	One has
	\begin{align}
	\text{rank}\Big(\begin{bmatrix}
	X_{0,N}\\U_{0,N}
	\end{bmatrix}\Big)=n + m,\label{condPE}
	\end{align}
	with $n$ and $m$ being the dimensions of state and input sets, respectively, if $U_{0,N}$ is a \emph{persistently exciting} input sequence of order $n+1$, i.e., $\text{rank}(U_{0,n+1,N})=m(n+1)$, where
	\begin{align*}
	U_{0,n+1,N}:=\begin{bmatrix}U_{0,N}(1)&	U_{0,N}(2)&\ldots&U_{0,N}(N-n)\\ U_{0,N}(2)&	U_{0,N}(3)&\ldots&U_{0,N}(N-n+1)\\ \vdots&\vdots&\ddots&\vdots\\U_{0,N}(n+1)&	U_{0,N}(n+2)&\ldots&U_{0,N}(N)\\\end{bmatrix}.
	\end{align*}
\end{lemma}

The condition of \emph{persistency of excitation} in Lemma~\ref{PEassumption} is common among direct data-driven approaches, since it ensures that the data in hand encode all information which is necessary for synthesizing controllers~\emph{directly} based on data~\cite{Willems2005note}.
Although Definition~\ref{opt3} and Theorem~\ref{thm:imp2} do not require this condition, the next result points out the difficulties in obtaining a feasible solution for $OP_d$, whenever condition~\eqref{condPE} does not hold.
\begin{corollary}\label{cor:PEreq}
	Consider the optimization problem $OP_d$ in Definition~\ref{opt3}, and the set
	\begin{align}
	\mathcal{F}:=\bigcap_{p=1}^{N}\mathcal{F}_p,\label{eq:Fins}
	\end{align}
	where $\mathcal{F}_p:=\Big\{(\tilde{A},\tilde{B})\in \mathbb{R}^{n\times n}\times \mathbb{R}^{n\times m}~\Big|~X_{1,N}(p) =\tilde{A}X_{0,N}(p)+\tilde{B}U_{0,N}(p)+d,d\in\Delta(\gamma)\Big\}$, in which $p\in[1,N]$.
	The set $\mathcal{F}$ is unbounded if and only if
	\begin{align}
	\text{rank}\Big(\begin{bmatrix}
	X_{0,N}\\U_{0,N}
	\end{bmatrix}\Big)<n + m.\label{cond}
	\end{align}
\end{corollary}

\begin{figure}
	\centering
	\includegraphics[width=9.5cm]{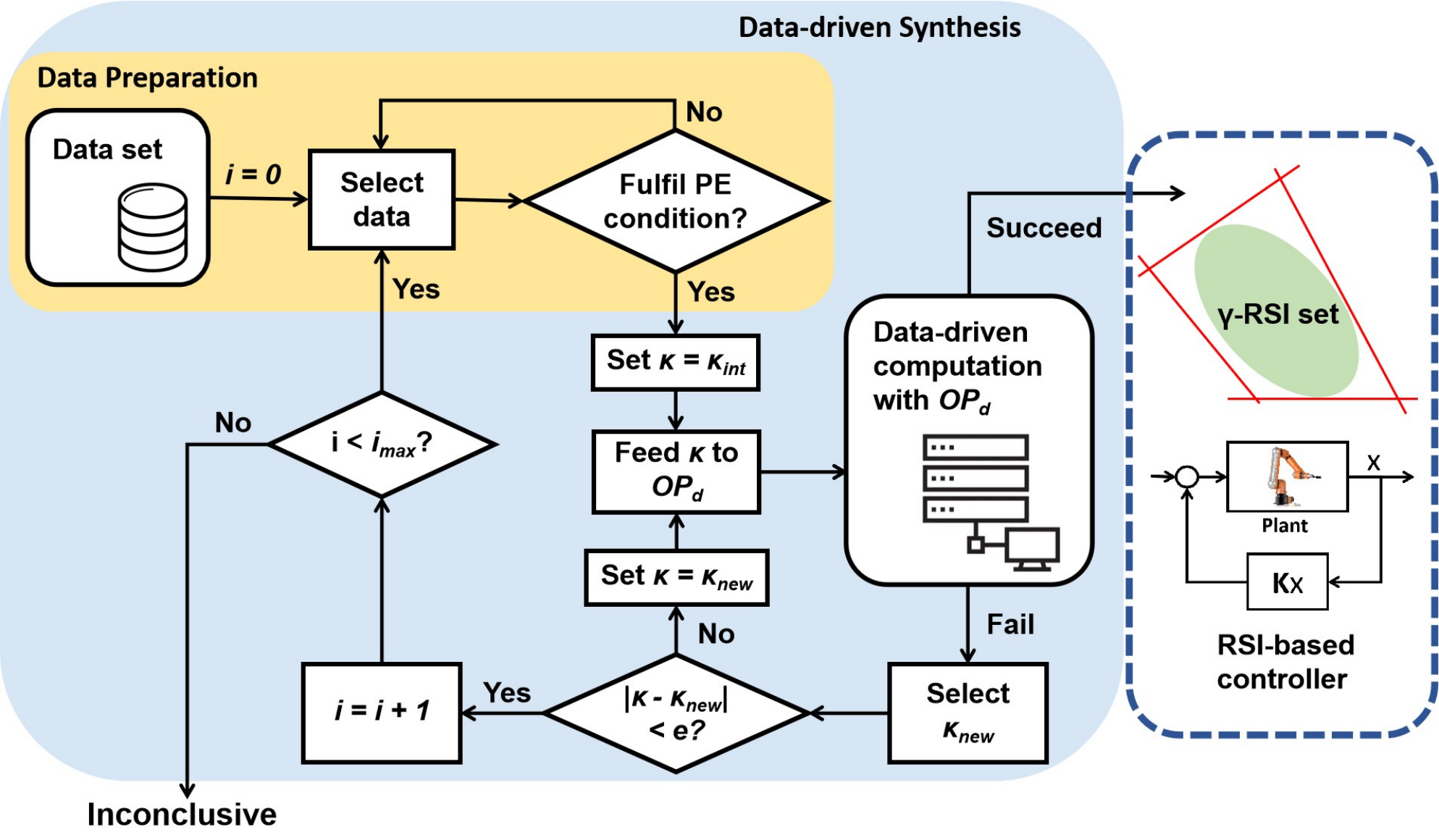}
	\caption{Flowchart of the proposed direct data-driven approach, with $OP_d$ and $\kappa$ as in Definition~\ref{opt3}, $\kappa_{int}\in(0,1]$, $e\in \mathbb{R}_{>0}$, and $i_{max}\in \mathbb{N}_{>0}$ being parameters which are manually selected by users, and PE condition referring to the condition of persistency of excitation as \\ in Lemma~\ref{PEassumption}.
	} \label{fig:alg_chart}
\end{figure}

The proof of Corollary~\ref{cor:PEreq} can be found in the Appendix.
As a key insight, given data of the form of~\eqref{eq:state_seq1} to~\eqref{eq:inputseqm}, the failure in fulfilling condition~\eqref{condPE} indicates that these data do not contain enough information about the underlying unknown system dynamics for solving the optimization problem $OP_d$, since the set of systems of the form of~\eqref{eq:linear_subsys} that can generate the same data is unbounded. 
Concretely,
the optimization problem $OP_d$ aims at finding a common $\gamma$-RSI set for any linear system as in~\eqref{eq:linear_subsys} such that $(A,B)\!\in\!\mathcal{F}$, with $\mathcal{F}$ as in~\eqref{eq:Fins}.
The unboundedness of the set $\mathcal{F}$ makes it very challenging to find a common $\gamma$-RSI set which works for all $(A,B)\!\in\!\mathcal{F}$.
In practice, to avoid the unboundedness of $\mathcal{F}$ and ensure that~\eqref{condPE} holds, one can increase the duration of the single input-state trajectory till the condition of persistency of excitation is fulfilled (cf. case studies).
Before proceeding with introducing the case study of this paper, we summarize in Figure~\ref{fig:alg_chart} a flowchart for applying the proposed direct data-driven approach.

\section{Case Studies}\label{sec4}
\begin{figure}
	\centering
	\includegraphics[width=3cm]{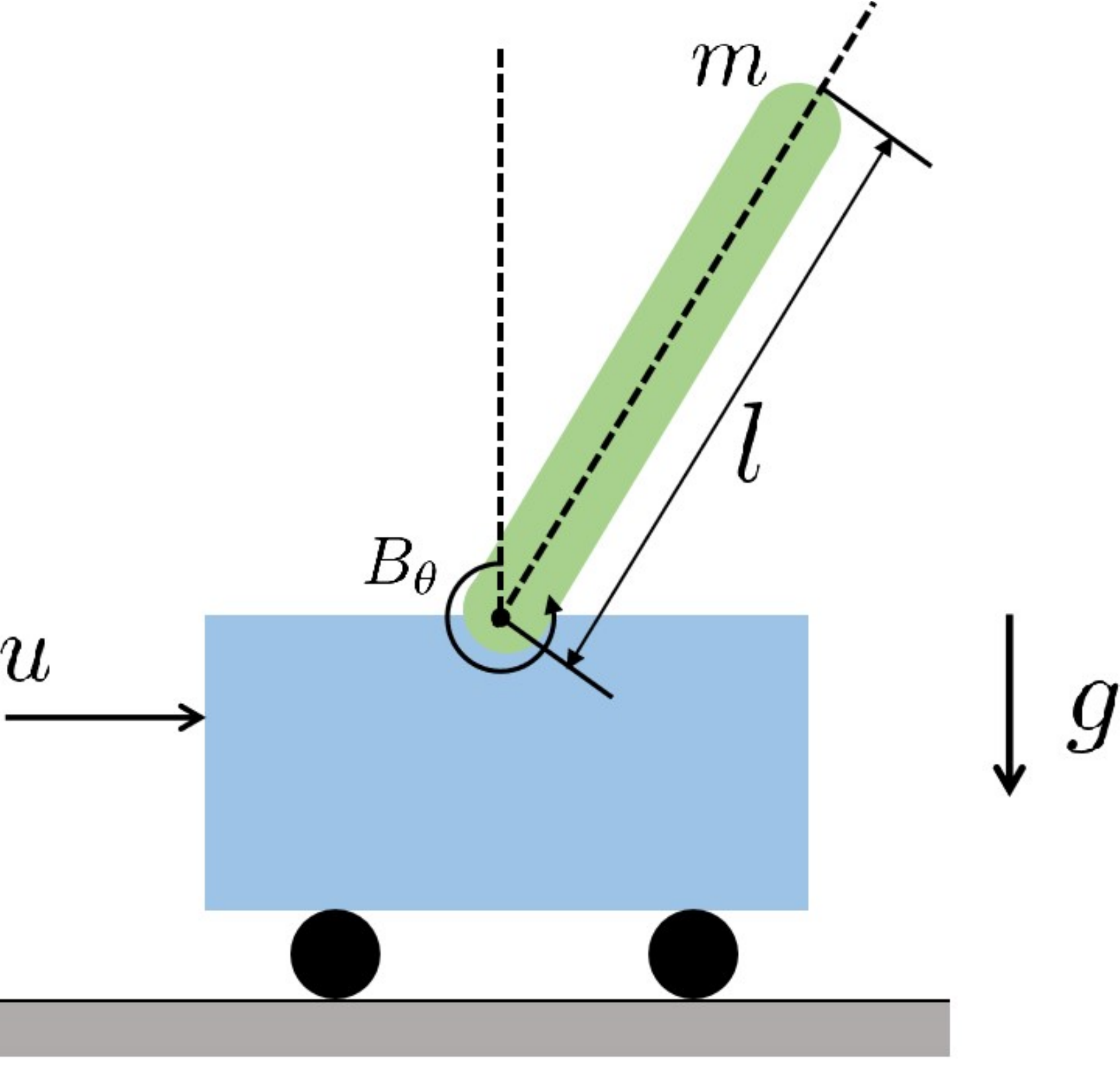}
	\caption{Inverted pendulum, where $m=0.1314\,$kg is the mass of the pendulum, $l=0.68\,$m is the length of the pendulum, $g=9.81\,$m/s is the gravitational constant, and $B_{\theta}=0.06\,$Nm/s is the damping coefficient of the connection between the cart (the blue part) and the pendulum (the green part).} \label{fig:IPS}
\end{figure}

To demonstrate the effectiveness of our results, we apply them to a four dimensional linearized model of the inverted pendulum as in Figure~\ref{fig:IPS}.
Although the direct data-driven approach proposed in Section~\ref{sec4.1} does not require any knowledge about matrices $A$ and $B$ of the model, we consider a model with known $A$ and $B$ in the case study mainly for collecting data, simulation, and computing the model-based gamma-RSI sets in Theorem~\ref{cor:synab} as baselines to evaluate the effectiveness of our direct data-driven approach (cf. Figure~\ref{fig1:IP12} and~\ref{fig1:IP34}).
When leveraging the direct data-driven method, we assume that $A$ and $B$ are fully unknown and treat the system as a black-box one.
The model of the inverted pendulum can be described by the difference equation as in~\eqref{eq:linear_subsys}, in which 
\begin{align}
A \!=\! \begin{small}\begin{bmatrix}1&0.02&0&0\\
0&1&0&0\\
0&0&1.0042&0.0194\\0&0&0.4208&0.9466\end{bmatrix}\end{small},
B \!=\! \begin{small}\begin{bmatrix}0.0002\\0.0200\\-0.0004\\[0.3em]-0.0429\end{bmatrix}\end{small},\label{dym}
\end{align}
where $x(k)=[x_1(k);x_2(k);x_3(k);x_4(k)]$ is the state of the system, with $x_1(k)$ being position of the cart, $x_2(k)$ being the velocity of the cart, $x_3(k)$ being the angular position of the pendulum with respect to the upward vertical axis, and $x_4(k)$ being the angular velocity of the pendulum;
$u(k)\in[-5,5]\ m/s^2$ is the acceleration of the cart that is used as the input to the system.
The safety objective for the inverted pendulum case study is to keep the position of the cart within $[-1,1]$ m, and the angular position of the pendulum within $[-\pi/12,\pi/12]$ rad.
This model is obtained by discretizing a continuous-time linearized model of the inverted pendulum as in Figure~\ref{fig:IPS} with a sampling time $\tau = 0.02s$, and including disturbances $d(k)$ that encompass unexpected interferences and model uncertainties.
The disturbances $d(k)$ belong to the set $\Delta(\gamma)$ as in~\eqref{eq:disturbance_set}, with $\gamma = (0.05\tau)^2$, which are generated based on 
a non-symmetric probability density function:
\begin{align}
f(d):=\left\{ 
\begin{aligned}
\frac{5}{\pi^2\gamma^2}&,\text{ for }d\in D_1;\\
\frac{9}{5\pi^2\gamma^2}&,\text{ for }d\in \Delta(\gamma)\backslash D_1,
\end{aligned}
\right.\label{noise1}
\end{align}
with $D_1:=\{ [d_1;d_2;d_3;d_4]\in  \Delta(\gamma)| d_i\in\mathbb{R}_{\geq 0}, i\in[1,4]\}$.
Here, we select the distribution as in~\eqref{noise1} to mainly illustrate the difficulties in identifying the underlying unknown system dynamics when the exogenous disturbances are subject to a non-symmetric distribution, even though they are bounded.
Meanwhile, our proposed direct data-driven approaches can handle such disturbances since we do not require any assumption on the disturbance distribution, e.g., being Gaussian or sub-Gaussian.
Moreover, this distribution is only used for collecting data and simulation, while the computation of data-driven $\gamma$-RSI sets does not require any knowledge of it.
The experiments are performed via MATLAB 2019b, on a machine with Windows 10 operating system (Intel(R) Xeon(R) E-2186G CPU (3.8 GHz)) and 32 GB of RAM.
The optimization problems in Section~\ref{sec3} are solved by using optimization toolboxes \texttt{YALMIP}~\cite{Lofberg2004YALMIP} and~\texttt{MOSEK}~\cite{ApS2019MOSEK}.

\begin{figure}
	\centering
	\includegraphics[width=7.3cm]{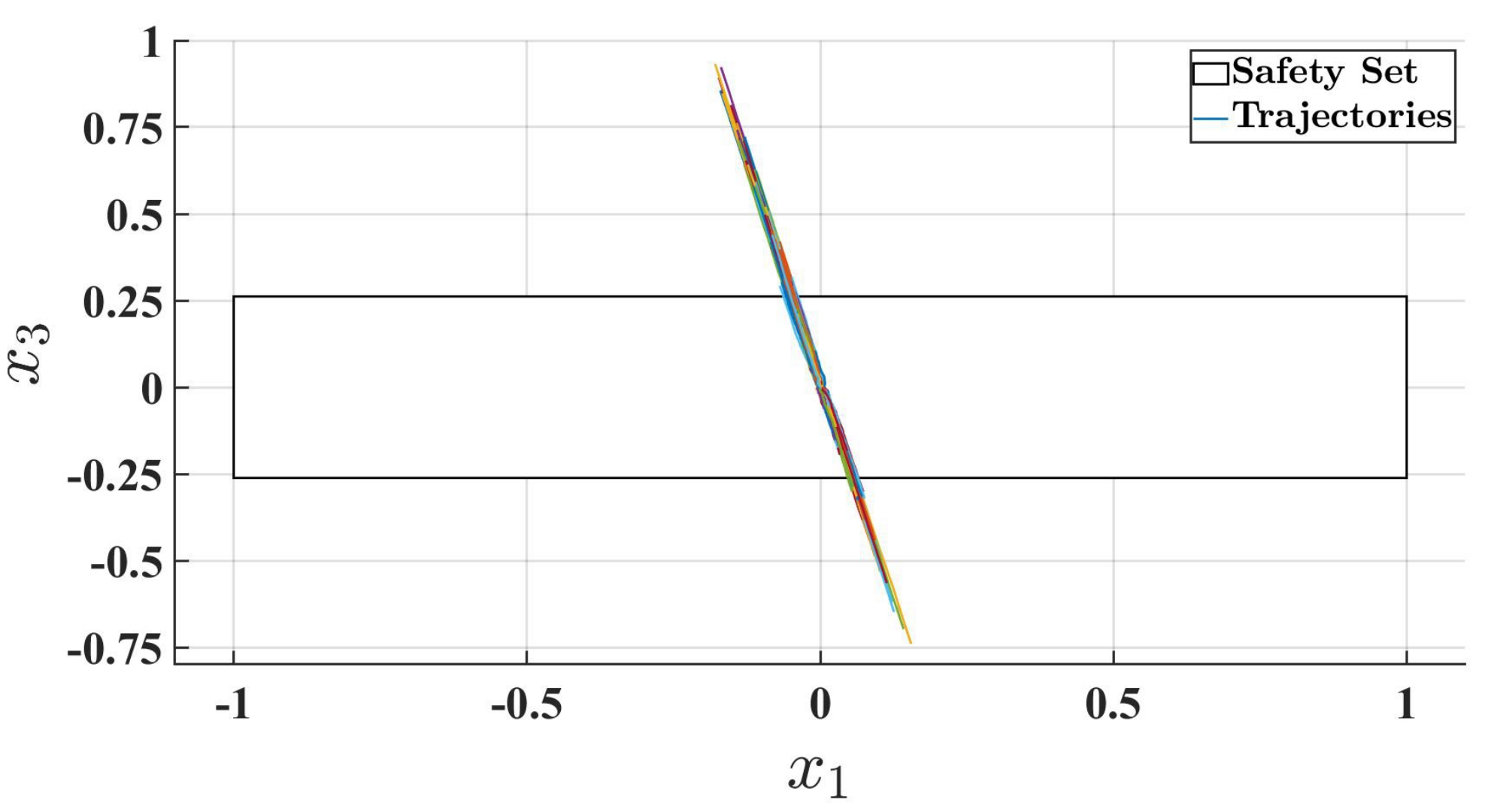}
	\caption{Projections of 1000 closed-loop trajectories on $x_1-x_3$ plane when applying the controller obtained by leveraging indirect data-driven approach.}\label{ind_traj}
\end{figure}

\begin{figure}
	\centering
	\includegraphics[width=7.3cm]{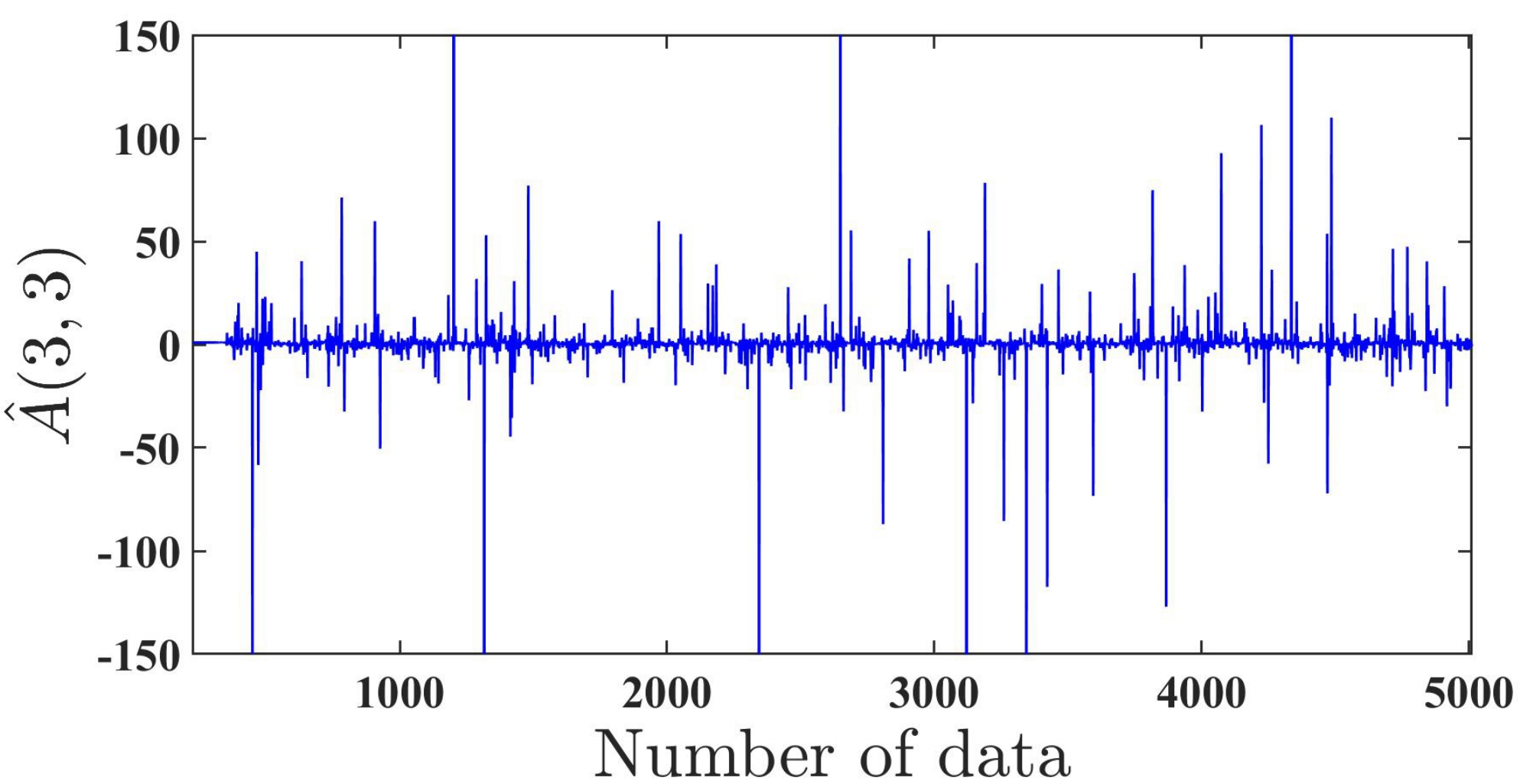}
	\caption{Evolution of the entry $\hat{A}(3,3)$ as number of data used for the system identification\\ increases.}\label{prob_ident}
\end{figure}

\begin{figure}
	\centering
	\includegraphics[width=7.3cm]{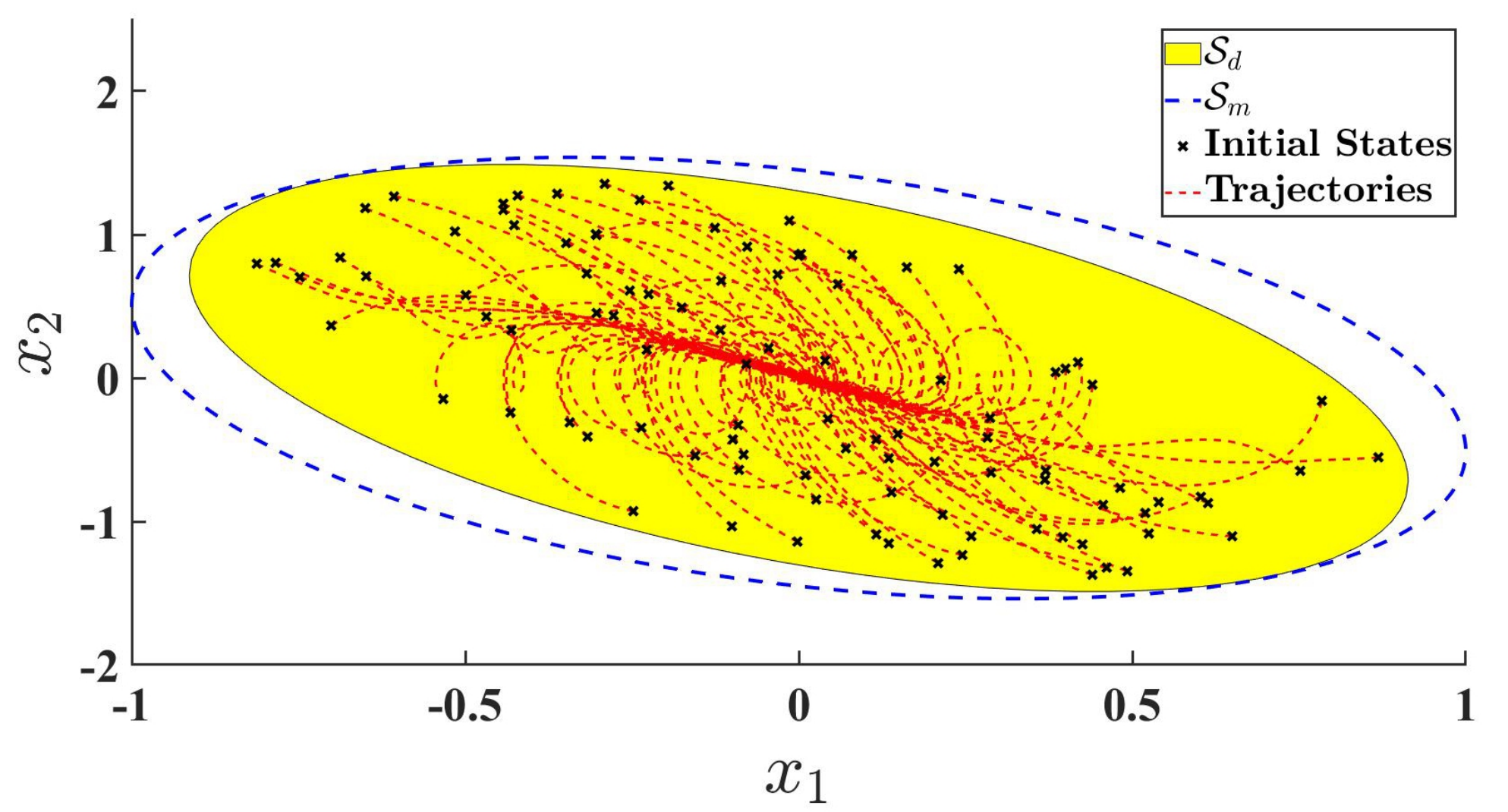}
	\caption{Projections of the data-driven $\gamma$-RSI set $\mathcal{S}_d$, the model-based $\gamma$-RSI set $\mathcal{S}_m$, initial states, and state trajectories on $x_1-x_2$ plane.}\label{fig1:IP12}
\end{figure}

\begin{figure}
	\centering
	\includegraphics[width=7.3cm]{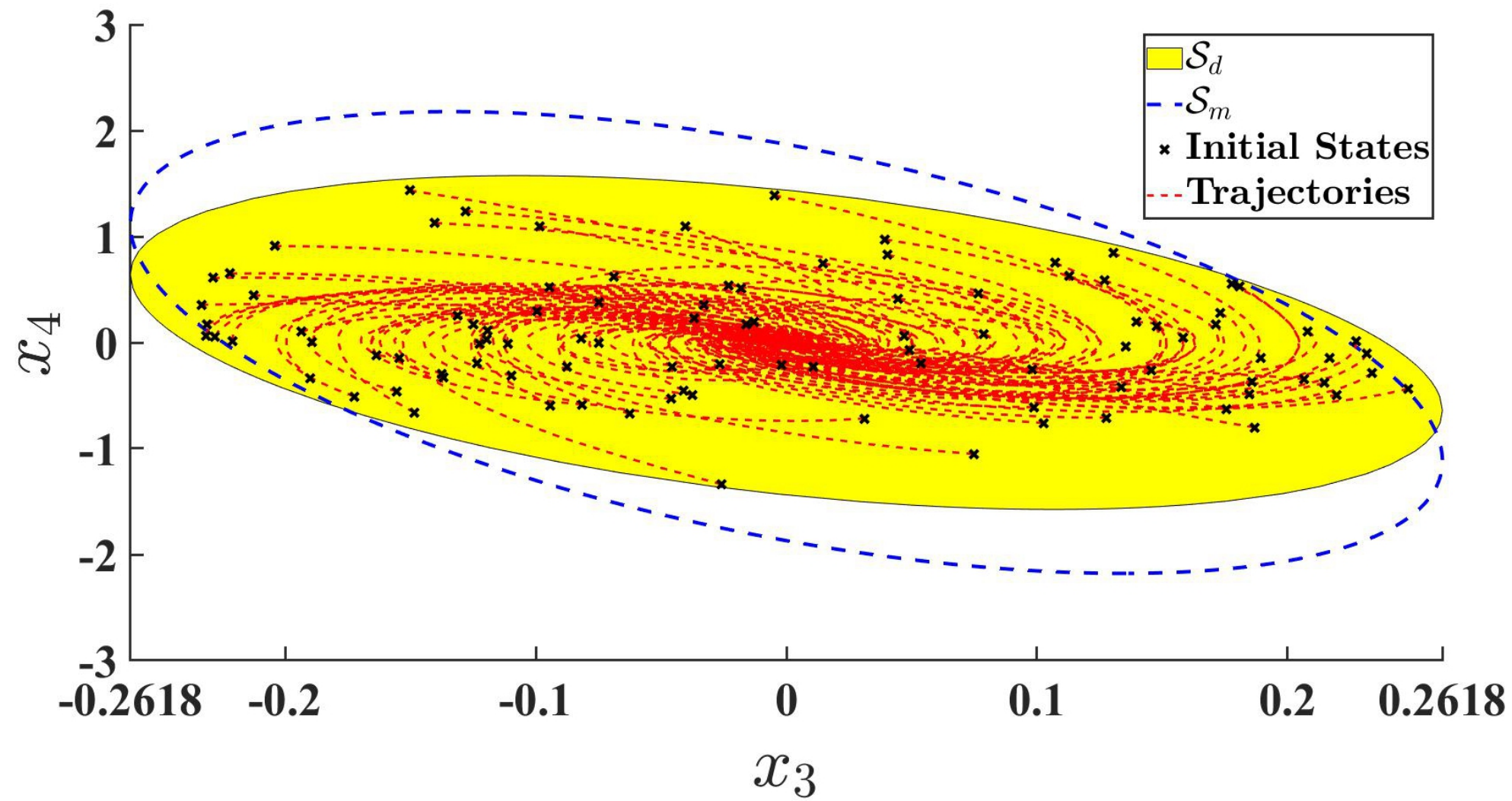}
	\caption{Projections of the data-driven $\gamma$-RSI set $\mathcal{S}_d$, the model-based $\gamma$-RSI set $\mathcal{S}_m$, initial states, and state trajectories on $x_3-x_4$ plane.}\label{fig1:IP34}
\end{figure}

First, we show the difficulties in applying indirect data driven approaches to solve Problem~\ref{prob} in our case study, when the bounded disturbances are generated based on a non-symmetric probability density function as in~\eqref{noise1}.
Here, we adopt least-squares approach as in~\cite{Jiang2004revisit} to identify matrices $A$ and $B$.
We collect data as in~\eqref{eq:state_seq1}-~\eqref{eq:inputseqm} with $N\!=\!500$, and we have the estimation of $A$ and $B$ as
\begin{align*}
\hat{A} \!=\! \begin{small}\begin{bmatrix}1&0.02&0&0\\
0&1&0&0\\
0.0764&-0.0888&2.3439&-0.3745\\0.0687&-0.0798&1.6255&0.5924\end{bmatrix}\end{small}\!\!,
\hat{B} \!=\! \begin{small}\begin{bmatrix}0.0002\\0.0200\\-0.0003\\[0.3em]-0.0422\end{bmatrix}\end{small}\!\!,
\end{align*}
respectively.
Based on the estimated model, we obtain a controller $u=K_ix$ by applying Theorem~\ref{cor:synab}, with $K_i= \big[-9.8089;-3.3176;$ $-112.7033;25.7470\big]^\top $.
With this controller, we initialize the system at $x =\big[0;0;0;0\big]^\top $, and simulate the system within the time horizon $H=70$.
The projections of closed-loop state trajectories on the $x_1-x_3$ plane are shown in Figure~\ref{ind_traj}, which indicate that the desired safety constraints are violated.
Additionally, we also depict in Figure~\ref{prob_ident} the evolution of the entry $\hat{A}(3,3)$ as an example to show that some of the entries in $\hat{A}$ keep fluctuating as the number of data used for system identification increases.
In other words, $\hat{A}$ does not seem to converge to the real value in~\eqref{dym} by increasing the number of data used for system identification.

Next, we proceed with demonstrating our direct data-driven approach.
To compute the data-driven $\gamma$-RSI set using Theorem~\ref{thm:imp2}, we first collect data as in~\eqref{eq:state_seq1}-\eqref{eq:inputseqm} with $N=107$.
Note that we pick $N=107$ such that condition~\eqref{condPE} holds.
Then, we obtain a data-driven $\gamma$-RSI set within $4.165$s.
Here, we denote the data-driven $\gamma$-RSI set by $\mathcal{S}_d := \{x\in\mathbb{R}^4 |x^\top P_dx\leq 1\}$, with
\begin{align*}
P_d = Q^{-1} =\begin{small}
\begin{bmatrix}3.3950 & 2.8786& 12.1264& 1.9861\\2.8786 & 3.8224 & 15.6826&2.7404\\12.1264&15.6826&81.9169&12.4079\\1.9861&2.7404&12.4079&2.4531\end{bmatrix}
\end{small},
\end{align*}
in which $Q$ is the solution of $OP_d$ with $\kappa=0.9813$.
The RSI-based controller associated with $\mathcal{S}_d$ is $u=K_dx$, where $K_d= \big[3.2672;4.9635;$ $38.1223;4.9989\big]^\top $.

\begin{figure}
	\centering
	\includegraphics[width=7.3cm]{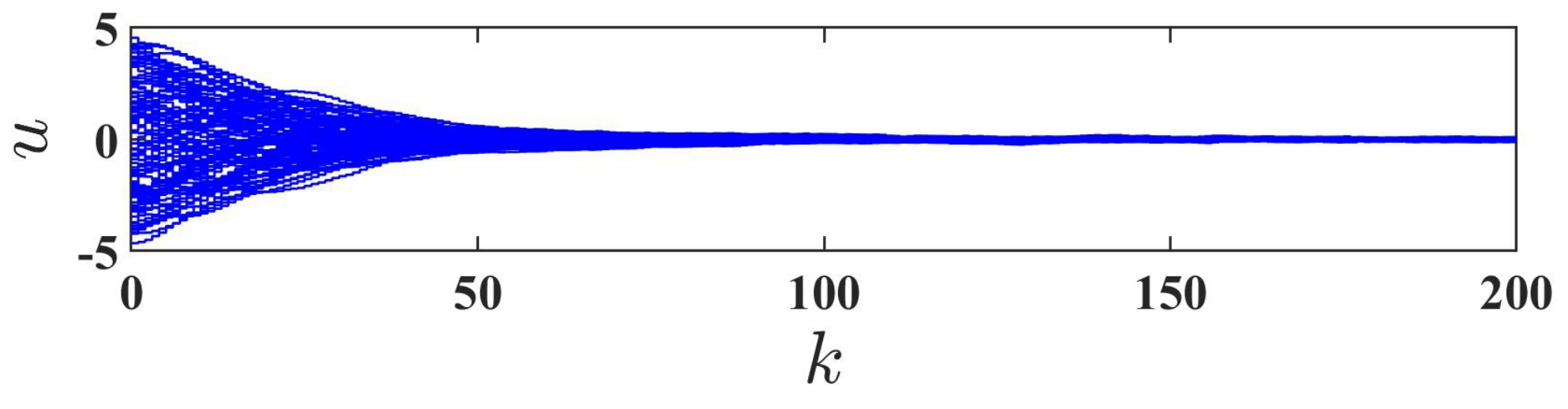}
	\caption{Input sequences for the inverted pendulum example.}\label{fig1:input_IP}
\end{figure}

As for the simulation, we first randomly select $100$ initial states from $\mathcal{S}_d$ following a uniform distribution.
Then, we apply the RSI-based controller associated with $\mathcal{S}_d$ in the closed-loop and simulate the system within the time horizon $H=200$.
In the simulation, disturbance at each time instant is injected following the distribution in~\eqref{noise1}.
The projections\footnote{Here, the projections of the $\gamma$-RSI sets are computed by leveraging \texttt{Ellipsoidal Toolbox}~\cite{Kurzhanskiy2021Ellipsoidal}.} of the data-driven $\gamma$-RSI sets, and closed-loop state trajectories on the $x_1-x_2$ and $x_3-x_4$ planes are shown in Figure~\ref{fig1:IP12} and~\ref{fig1:IP34}, respectively.
For comparison, we also compute the model-based $\gamma$-RSI set with Theorem~\ref{cor:synab}, denoted by $\mathcal{S}_m$, and project it onto relevant coordinates.
One can readily verify that all trajectories are within the desired safety set, and input constraints are also respected, as displayed in Figure~\ref{fig1:input_IP}.
It is also worth noting that, as shown in Figure~\ref{fig1:IP34}, the data-driven $\gamma$-RSI set does not necessarily need to be inside the model-based one, since the $\gamma$-RSI set with the maximal volume (cf. Remark~\ref{objective}) do not necessarily contain all other possible $\gamma$-RSI sets with smaller volume. 

\section{Conclusions}\label{sec5}
In this paper, we proposed a direct data-driven approach to synthesize safety controllers, which enforce invariance properties over unknown linear systems affected by unknown-but-bounded disturbances.
To do so, we proposed a direct data-driven framework to compute $\gamma$-robust safety invariant ($\gamma$-RSI) sets, which is the main contribution of this paper. 
Moreover, we discuss the relation between our proposed data-driven approach and the condition of persistency of excitation, explaining the difficulties in finding a suitable solution when the collected data do not fulfill such a condition.
To show the effectiveness of our results, we apply them to a 4-dimensional inverted pendulum.
Providing a data-driven approach for computing control barrier functions to enforce invariance properties is under investigation, as a future work.

\section{Acknowledgment}
This work was supported in part by the H2020 ERC Starting Grant AutoCPS (grant agreement No 804639) and by an Alexander von Humboldt Professorship endowed by the German Federal Ministry of Education and Research.


\bibliographystyle{my-elsarticle-num}    
\bibliography{sample-base}

\section*{Appendix}
\renewcommand{\theequation}{A.\arabic{equation}}
{\bf Proof of Theorem~\ref{thm:LMI_1}}: 
We first show that the statement regarding \textit{if} holds.	
If $\exists \kappa \in (0,1]$ such that $(y+\tilde{d})^\top P(y+\tilde{d})\leq 1$ holds $\forall y\in \mathbb{R}^n$ with $y^\top Py\leq \kappa$, and  $\forall \tilde{d}$ with $\tilde{d}^\top \tilde{d}\leq \gamma$, one can let $y=(A+BK)x$ with $x^\top Px\leq 1$ without loss of generality. 
This immediately implies that~\eqref{ineq:safety_set} holds $\forall x\in \mathbb{R}^n$ and $\forall d\in\mathbb{R}^n$ with $d^\top d\leq \gamma$.

Next, we show that the statement regarding \textit{only if} also holds by contradiction.
Suppose that $\nexists \kappa \in(0,1]$ such that (\textit{\textbf{Cond.1}}) holds. 
Then, $\exists x\in\mathbb{R}^n$, with $x^\top Px\leq 1$, such that $x^\top (A+BK)^\top P(A+BK)x>1$. 
Accordingly, one has $\Big((A+BK)x+d\Big)^\top P\Big((A+BK)x+d\Big)> 1$ with $d = \mathbf{0}_n$, which results in a contradiction to the fact that~\eqref{ineq:safety_set} holds for $\forall d\in\Delta(\gamma)$.
Therefore, one can see that there exists $\kappa \in(0,1]$ such that (\textit{\textbf{Cond.1}}) holds if~\eqref{ineq:safety_set} holds $\forall d\in\Delta(\gamma)$, and $\forall x\in \mathbb{R}^n$, with $ x^\top Px\leq 1$.
In the following discussion, we denote such $\kappa$ by $\kappa'$.
Similarly, assuming that $\nexists \kappa \in(0,1]$ such that (\textit{\textbf{Cond.2}}) holds. 
This indicates that $\forall \kappa \in(0,1]$, $\exists y\in\mathbb{R}^n$, with $y^\top Py\leq \kappa$, or $\exists \tilde{d}\in\Delta(\gamma)$ such that $(y+\tilde{d})^\top P(y+\tilde{d})> 1$.
Let's consider $\kappa = \kappa'$ and we can let $y=(A+BK)x$ with $x^\top Px\leq 1$ without loss of generality.
Then,  $\exists x\in\mathbb{R}^n$, with $x^\top Px\leq 1$, or $\exists d\in\Delta(\gamma)$, such that $\Big((A+BK)x+d\Big)^\top P\Big((A+BK)x+d\Big)> 1$, which is contradictory to the fact that~\eqref{ineq:safety_set} holds $\forall d\in\Delta(\gamma)$, and $\forall x\in \mathbb{R}^n$ with $ x^\top Px\leq 1$.
Hence, there exists $\kappa \in(0,1]$ s.t. (\textit{\textbf{Cond.2}}) holds, if~\eqref{ineq:safety_set} holds $\forall d\in\Delta(\gamma)$, and $\forall x\in \mathbb{R}^n$, with $ x^\top Px\leq 1$, which completes the proof. $\hfill\blacksquare$

{\bf Proof of Theorem~\ref{cor:synab}}: 
First, we show that given a $\kappa \in(0,1]$, (\textit{\textbf{Cond.1}}) in Theorem~\ref{thm:LMI_1} holds if and only if~\eqref{synab_cond1} holds.
By applying S-procedure~\cite[Section B.2]{Boyd2004Convex}, (\textit{\textbf{Cond.1}}) in Theorem~\ref{thm:LMI_1} holds if and only if there exists $\lambda\in\mathbb{R}_{\geq 0}$ s.t.
\begin{align}
\begin{bmatrix}(A+BK)^\top P(A+BK) & \mathbf{0}\\\mathbf{0} &-\kappa\end{bmatrix}\preceq \lambda \begin{bmatrix}P &\mathbf{0}\\\mathbf{0} &-1\end{bmatrix}\label{eq:help1},
\end{align}
holds.
Accordingly,~\eqref{eq:help1} holds if and only if $(A+BK)^\top P(A+BK)\preceq \lambda P$ with $\lambda \leq \kappa$.
Hence,~\eqref{eq:help1} holds if and only if~\eqref{synab_cond1} holds according to~\cite[Theorem 1.12]{Zhang2006Schur}, with $Q = P^{-1}$.

Next, we proceed with showing that (\textit{\textbf{Cond.2}}) in Theorem~\ref{thm:LMI_1} holds if and only if~\eqref{synab_cond2} holds.
First, considering the geometric properties of ellipsoids $x^\top Px\leq 1$ and $x^\top Px \leq \kappa$, the shortest distance between both ellipsoids is $\sqrt{\lambda_{min}}-\sqrt{\kappa\lambda_{min}}$, with $\lambda_{min}$ the minimal eigenvalue of $P^{-1}$.
Hence, to ensure (\textit{\textbf{Cond.2}}), we need to guarantee that 
\begin{equation}
\sqrt{\lambda_{min}}-\sqrt{\kappa\lambda_{min}}\geq \sqrt{\gamma}.\label{mid_ineq}
\end{equation}
Accordingly, 
\begin{itemize}
	\item If $\kappa\neq 1$, ~\eqref{mid_ineq} requires that $\lambda_{min}\geq \frac{\gamma}{(1-\sqrt{\kappa})^2}$, which holds if and only if $P^{-1}\succeq \frac{\gamma}{(1-\sqrt{\kappa})^2} \mathbf{I}$;
	\item If $\kappa = 1$,~\eqref{mid_ineq} holds if and only if $\gamma = 0$, for any $\lambda_{min}\geq 0$.
	Hence,~\eqref{mid_ineq} holds if and only if $P^{-1}\succeq 0$.
\end{itemize}
Therefore, (\textit{\textbf{Cond.2}}) in Theorem~\ref{thm:LMI_1} holds if and only if~\eqref{synab_cond2} holds.

Finally, we show that~\eqref{synab_cond3} and~\eqref{synab_cond4} are respecting the safety set as in~\eqref{safety_set}, and input constraints as in~\eqref{input_set}, respectively.
According to~\cite[Lemma 4.1]{Seto1999Engineering}, ~\eqref{eq:safety_set} holds for $S$ as in~\eqref{safety_set} if and only if~\eqref{synab_cond3} holds.
Similarly, considering the the RSI-based controller as in~\eqref{eq:safety_controller},~\eqref{input_set} requires that $b_jKx\leq 1$ should hold for all $j = 1,\ldots,\mathsf{j}$.
This can be enforced by~\eqref{synab_cond4} according to~\cite[Lemma 4.1]{Seto1999Engineering} and~\cite[Theorem 1.12]{Zhang2006Schur}, which completes the proof. $\hfill\blacksquare$

{\bf Proof of Theorem~\ref{thm:imp2}}: 
One can verify that~\eqref{LMI2_cond1}, ~\eqref{LMI2_cond3}, and~\eqref{LMI2_cond4}, are the same as~\eqref{synab_cond2},~\eqref{synab_cond3}, and~\eqref{synab_cond4}, respectively.
Therefore, we show that~\eqref{LMI2_cond2}, with $\epsilon_i\!>\!0$, $\forall i\!\in\![1,N]$, implies~\eqref{synab_cond1} in the remainder of this proof.
According to~\cite[Theorem 1.12]{Zhang2006Schur},~\eqref{synab_cond1} holds if and only if $(A+BK)Q(A+BK)^\top \preceq \kappa Q$, when considering the Schur complement of $\kappa Q$ of the matrix on the left hand side of~\eqref{synab_cond1}, with $K=\bar{K}Q^{-1}$.
Therefore, ~\eqref{synab_cond1} holds if and only if
\begin{align}
\begin{bmatrix}\mathbf{I} & A &B\end{bmatrix} \begin{bmatrix}\kappa Q & \mathbf{0} & \mathbf{0}\\\mathbf{0} &-Q& -\bar{Z}^\top \\ \mathbf{0}& -\bar{Z}& -KQK^\top \end{bmatrix}\begin{bmatrix}\mathbf{I} \\A^\top  \\B^\top \end{bmatrix}\succeq 0, \label{ineq:matrixQ}
\end{align}
holds, with $\bar{Z}= KQ$. 
Next,  we show that~\eqref{LMI2_cond2}, with $\epsilon_i>0$, $\forall i\in[1,N]$, implies~\eqref{ineq:matrixQ} holds for any $A\!\in\!\mathbb{R}^{n\times n}$ and $B\!\in\!\mathbb{R}^{n\times m}$ such as $X_{1,N} = AX_{0,N}+BU_{0,N} +D$ holds with $D = [d(0)\ldots d(k)\ldots$ $d(N-1)]$ and $d(k)^\top d(k)\leq \gamma$, $\forall k\in[0,N-1]$, indicating that~\eqref{ineq:matrixQ} holds for the unknown $A$ and $B$ as in~\eqref{eq:linear_subsys}.
Considering~\eqref{eq:linear_subsys}, since $d(k)d(k)^\top \!\!\!\preceq \gamma \mathbf{I}$, $\forall d(k)\in\Delta(\gamma)$, one has
\begin{align}
\begin{bmatrix}\mathbf{I} & A &B\end{bmatrix}\bar{N}_p
\begin{bmatrix}
\gamma \mathbf{I}    &\mathbf{0} \\
\mathbf{0}   &-1
\end{bmatrix}
\bar{N}_p^\top \begin{bmatrix}\mathbf{I} & A &B\end{bmatrix}^\top \succeq 0,\label{matrixS}
\end{align}
$\forall p\in[1,N]$, with 
\begin{align}
\bar{N}_p:= \begin{bmatrix}
\mathbf{I}  &X_{1,N}(p) \\
\mathbf{0}  &-X_{0,N}(p)  \\
\mathbf{0}	&-U_{0,N}(p)  
\end{bmatrix}.\label{Np}
\end{align}
Considering~\cite[Theorem 1.12]{Zhang2006Schur}, if $\exists \epsilon_i>0$, $\forall i\in[1,N]$ such that~\eqref{LMI2_cond2} holds, then one gets
\begin{small}
	\begin{align}
	\begin{bmatrix}\kappa Q\! & \mathbf{0}\! & \mathbf{0}\\\mathbf{0}\! &-Q\!& -\bar{Z}^\top \\ \mathbf{0\!}& -\bar{Z}\!& 0\end{bmatrix}\!\! -\!\!\begin{bmatrix}\mathbf{0} \\\mathbf{0} \\\bar{Z}\end{bmatrix}Q^{-1}\!\!\begin{bmatrix}\mathbf{0} \\\mathbf{0} \\\bar{Z}\end{bmatrix}^\top \!\!\!\!\!-\! \! \!\sum_{p=1}^{N}\epsilon_p \bar{N}_p
	\!\!\begin{bmatrix}
	\gamma \mathbf{I}    &\mathbf{0} \\
	\mathbf{0}   &-1
	\end{bmatrix} \!
	\bar{N}_p^\top \!\!\succeq\!\! 0, \label{eq:int}
	\end{align}
\end{small}

\noindent
with $\bar{N}_p$ as in~\eqref{Np}.
According to~\cite[Lemma 2]{Bisoffi2021Trade},~\eqref{eq:int} implies that~\eqref{ineq:matrixQ} holds for all~\eqref{matrixS} with $p\in[1,N]$, which completes the proof. $\hfill\blacksquare$

{\bf Proof of Corollary~\ref{cor:PEreq}}: 
Consider the system as in~\eqref{eq:linear_subsys}, $X_{1,N}$ and $X_{0,N}$, and $U_{0,N}$ as in~\eqref{eq:state_seq1} to~\eqref{eq:inputseqm}, respectively.
Given any disturbance sequence $D:=[d_1\ d_2\ \ldots\ d_N]\in(\Delta(\gamma))^N$, with $(\Delta(\gamma))^N$ being the Cartesian product of $N$ times of the set $\Delta(\gamma)$, we define	 $\tilde{X}_D:=(X_{1,N}-D)^\top \!$, and $Y_{0,N} := \begin{bmatrix}X^\top _{0,N}\  U^\top _{0,N}\end{bmatrix}$.
Then, by definition of the set $\mathcal{F}$ as in~\eqref{eq:Fins}, one has
\begin{align}
\mathcal{F}=\bigcup_{D\in(\Delta(\gamma))^N}\mathsf{W}(D)^\top ,\label{eq:toprove}
\end{align}
with $\mathsf{W}(D):=\{W\in\mathbb{R}^{(n+m)\times n}|\tilde{X}_D= Y_{0,N}W,D\in(\Delta(\gamma))^N\}$.

Firstly, we show that the statement regarding \textit{if} holds.
To this end, we first show that the set $\mathsf{W}(D)$ is either unbounded or empty, when~\eqref{cond} holds.
Consider the equation $\tilde{X}_D= Y_{0,N}W$, in which $W\in\mathbb{R}^{(n+m)\times n}$ is an unknown matrix to be determined (note that there may not be suitable $W$, the discussion comes later).
According to~\cite[Section 3.3]{Strang2016Introduction}, for any column $W(i)$, $i\in[1,n]$, if there exists
\begin{align}
{}^i\alpha:=\begin{bmatrix}{}^i\alpha_1&{}^i\alpha_{2}&\ldots&{}^i\alpha_{n+m}\end{bmatrix}\in\mathbb{R}^{n+m},\label{eqhelp1}
\end{align}
such that 
\begin{align}
\tilde{X}_D(i)=\sum_{\mathsf{a}=1}^{n+m}{}^i\alpha_{\mathsf{a}}Y_{0,N}(\mathsf{a})\label{eqhelp2}
\end{align}
holds, then one has $W(i)\in \{{}^i\alpha + w~|~ w\in \mathsf{ker}(Y_{0,N})\}$, with $\mathsf{ker}(Y_{0,N})$ the kernel of $Y_{0,N}$; otherwise, one has $W(i)\in \emptyset$.
Therefore, one has
\begin{align}
\mathsf{W}(D)=\prod_{i=1}^{n}\{{}^i\alpha + w~|~ w\in \mathsf{ker}(Y_{0,N})\}\neq \emptyset ,\label{eq:WD}
\end{align}
when for all $i\in[1,n]$, there exists ${}^i\alpha$ as in~\eqref{eqhelp1} such that~\eqref{eqhelp2} holds; and $\mathsf{W}(D)=\emptyset$ otherwise.
Note that $\mathsf{ker}(Y_{0,N})$ is an $\mathsf{r}$-dimension subspace of $\mathbb{R}^{n+m}$, with $\mathsf{r}=n+m-\text{rank}(Y_{0,N})$ according to~\cite[Section 3.5]{Strang2016Introduction}.
If~\eqref{cond} holds, then one has $\mathsf{r}>0$.
In this case, the set $\{{}^i\alpha + w~|~ w\in \mathsf{ker}(Y_{0,N})\}$ is unbounded for any ${}^i\alpha\in\mathbb{R}^{n+m}$ due to the unboundedness of $\mathsf{ker}(Y_{0,N})$.
As a result, the set $\mathsf{W}(D)$ is either unbounded or empty when~\eqref{cond} holds.
Moreover, since $X_{0,N}$, $X_{1,N}$, and $U_{0,N}$ are data collected from the system as in~\eqref{eq:linear_subsys}, we always have $[A\ B]\in \mathsf{W}(D)$ for some $D\in\Delta(\gamma))^N$, with $A$ and $B$ the unknown matrices in~\eqref{eq:linear_subsys}.
In other words, there always exists $D\in\Delta(\gamma))^N$ such that $\mathsf{W}(D)$ is not empty (and is therefore unbounded).
Hence, it is then straightforward that the right-hand side of~\eqref{eq:toprove} is unbounded, so that statement regarding \textit{if} holds.

Next, we show that the statement regarding \textit{only if} also holds by showing $\mathcal{F}$ is bounded when 
\begin{align}
\text{rank}\Big(\begin{bmatrix}X_{0,N}\\U_{0,N}\end{bmatrix}\Big)=n + m.\label{pe}
\end{align}
When~\eqref{pe} holds, then $\mathsf{ker}(Y_{0,N})$ only contains the origin.  
As a result, the set $\{{}^i\alpha + w~|~ w\in \mathsf{ker}(Y_{0,N})\}$ is either a singleton set that only contains ${}^i\alpha$, or an empty set, so that the set $\mathsf{W}(D)$ is either a singleton set or an empty empty set, when~\eqref{pe} holds.
Then, the boundedness the right-hand side of~\eqref{eq:toprove} follows by the boundedness of the set $\Delta(\gamma))^N$, which completes the proof.$\hfill\blacksquare$

\end{document}